\documentstyle[aps,preprint]{revtex}

\author{Diego F. Torres\thanks{e-mail: dtorres@venus.fisica.unlp.edu.ar}}
 
\address{Departamento de F\'{\i}sica - Universidad Nacional de La Plata\\
C.C. 67, C. P. 1900, La Plata, Buenos Aires, Argentina}

\title{Classes of Anisotropic Cosmologies of Scalar-Tensor Gravitation}

\begin{document}

\maketitle

\begin{abstract}

A study of an algorithm method capable to reveal
anisotropic solutions of general scalar-tensor gravitation
--including non-minimally couplings-- is presented. 
It is found that it is possible to classify the 
behavior of the field of different scalar-tensor theories
in equivalence classes, with the same classifier function
that was obtained in Friedmann-Robertson-Walker models.

\noindent PACS {\it number(s)}: 04.50.+h, 04.20.Cv, 98.80.Cq, 98.80.Hw

\end{abstract}

\newpage

Scalar-tensor gravitation have proved to be a useful tool in the understanding
of early universe models. The first and best known case of such theories
is Brans-Dicke (BD) gravity \cite{BD}, in which there is a coupling function
$w(\phi)$ equal to a constant; $\phi$ being a dynamical field related with
the previous gravitational constant.
More general theories with other couplings
have also been studied \cite{STG}. The interest on these theories have been
recently rekindled by inflationary scenarios \cite{INF} and fundamental theories
which seek to incorporate gravity with other forces of nature. Particularly,
in string theories, a dilaton field coupled to curvature appears in the low energy
effective action \cite{strings}. When scalar-tensor gravitation is concerned, one is
interested also in the cosmological models it leads. Observational constraints,
mainly coming from the weak field tests \cite{WILL} and nucleosynthesis \cite{Nucleo},
put several bounds upon the couplings. In any case, in order to evaluate
the cosmological scenario and to test the predictable force of any theory,
it is desirable to have 
exact analytical solutions of the field equations. 
But it was not until a few years
ago, that Barrow \cite{Barrow1}, 
Barrow and Mimoso \cite{Barrow2} and Mimoso
and Wands \cite{MW} derived algebraic-numerical
methods that allow
exact Friedmann-Robertson-Walker (FRW) 
solutions to be found 
in models with matter content in the form of a barotropic fluid 
for any kind of coupling $w(\phi)$. 

However, scalar-tensor theories can be formulated in two 
different ways depending on
the choice of the basic action.
These two possible choices are the BD one, in which there is an arbitrary
function in the kinetic term of the scalar, and the one which admit
an arbitrary function multiplying the curvature while mantaining a 
common kinetic term,
for the theories known as non-minimally coupled (NMC). 
Via a field redefinition
one can stablish the equivalence between these choices 
in the most of the cases,
but that is not what happens when the functions involved are not analytically
invertible as, for instance, in the hyperextended inflationary scenario
\cite{LiddleWands}. Very recently, we have presented a study 
on the full lagrangian
density for the field, which involves, 
in the more general case, two free functions
\cite{TORRES_V}.
This lagrangian reads

\begin{equation}
\label{HSTG}L=16\pi L_m-\frac{w(\phi )}{\phi}
\phi _{,\mu }
\phi ^{,\mu }+G(\phi
)^{-1}R
\end{equation}
where, as usual, $L_m$ refers to the matter lagrangian density and $R$
is the curvature scalar.
Each possible choice of the action
may be reproduced by a convenient selection
of $G$ and/or $w$.
We have called 
hyperextended scalar-tensor gravitation (HSTG) to
the theories of gravity this lagrangian leads, 
because of the similarity with the inflationary model.
Similar algorithm methods of 
massless
scalar fields, developed by Mimoso and Wands in \cite{MW},
were applied to this general approach.
Those methods 
allowed
to compute exact analytical solutions 
for FRW models in vacuum or with matter
content consisting in radiation or stiff fluids 
for any choice of $G$ and $w$
simultaneously \cite{TORRES_V}. That includes the cases of NMC, 
where solutions are scarce
\cite{NMCO}. As spinoff of that research, we find 
that it was possible to define an
equivalence classes behavior of scalar-tensor 
gravitation in which the scalar field
itself is a class variable. 
Specifically, for different lagrangian densities
--a set $(G,w)$-- that have an equal function of $\phi$ given by

\begin{equation}
\label{alpha}\alpha =\left( \frac \phi G\right) ^2\left( \frac{dG}{d\phi }%
\right) ^2+\frac 23w G\phi 
\end{equation}
we find the same solution for the field as a function of the conformal time
\footnote{ The conformal time is defined as $d\eta=a dt$ where $a$ is the scale
factor}\cite{TORRES_V}. 
Then, inside each class, the scale factor could be obtained inmediately
as $a^2=G X$ where $X$ was a class-dependent function of time. 
This approach was
also used in the search of slow-roll solutions for non-minimally coupled
theories \cite{TORRES_S}.

Anisotropic homogeneous cosmological models are being intensively
studied since quite a long time \cite{MAHM}. In particular, within scalar-tensor
gravitation, the analysis of anisotropic cosmologies could reveal different behavior
when compared with Einstein General Relativity 
near the singularity \cite{RF} or in the inflationary epoch \cite{PIMENTEL}.
Processes of isotropization of Brans-Dicke Bianchi-type solutions are also of
current interest \cite{Chauvet}.

The aim of this work  
is to explore, within the general lagrangian density
of HSTG, the case of some anisotropic cosmological models. In particular, we
are interested in examine if is it possible to extend previously derived
results for anisotropic models in BD gravity to this general approach;
allowing, for instance, 
to compute also anisotropic solutions for non-minimally
coupled theories. In addition, we want to observe if anisotropy do or do not
break the classification scheme with $\alpha$ 
as the classifier
function. Throughout we draw heavily on the results on massless fields in
anisotropic universes \cite{MWa} and in the properties of the lagrangian
(\ref{HSTG}) when considered in the Einstein frame.

\vspace{1.2 cm}

The field equations of HSTG are 
\footnote{ Equation (\ref{movimiento}) is obtained 
after the assumption that $\phi$ depends only on time.}

\begin{equation}
\label{campo}R_{\mu \nu }-\frac 12g_{\mu \nu }R=G(\phi )\left[ 8\pi T_{\mu \nu
}+\frac \omega \phi \phi _{,\mu }\,\phi ^{,\mu }-\frac \omega {2\phi }\phi
_{,\alpha }\,\phi ^{,\alpha }g_{\mu \nu }+(G^{-1})_{,\mu
;\nu }-g_{\mu \nu }\Box (G^{-1})\right]
\end{equation}

\begin{equation}
\label{movimiento}\dot \phi ^2\left[ \frac 1\phi \frac{d\omega }{d\phi }-\frac
\omega {\phi ^2}+G\ \frac{dG^{-1}}{d\phi }\frac \omega \phi \right] +\frac{%
2\omega }\phi \ \Box \phi +3G\ \frac{dG^{-1}}{d\phi }\Box (G^{-1})-G\ \frac{%
dG^{-1}}{d\phi }8\pi T=0
\end{equation}
It is very important to remark that the usual relation $T_{\quad ;\nu }^{\mu
\nu }=0$ stablishing the conservation laws (in the meaning of GR) of the
matter fields holds true. 
Note that when $G=1/\phi$, these equations reduce to the common BD ones. 
The Einstein
frame was introduced by Dicke \cite{Dicke}, when working in 
Brans-Dicke gravity, by defining a
conformal transformation in the form

\begin{equation}
\tilde g_{ab}=G_0\phi g_{ab}
\end{equation}
where $G_0$ is an arbitrary constant which becomes the gravitational constant
in the transformed frame. In a similar fashion, we introduce 

\begin{equation}
\tilde g_{ab}=G_0 G(\phi)^{-1} g_{ab}
\end{equation}
Using the relation between the curvature scalars of the common --Jordan-- frame
and the transformed --Einstein-- frame, given for all conformal transformations
by Synge \cite{Synge}, we get for the action

\begin{equation}
S_{EF}= \frac{1}{16 \pi} \int \sqrt{- \tilde g} \left[ \frac{1}{G_0} \tilde R -
\tilde g^{ab} \frac{ \phi_{,a} \phi_{,b}}{\phi^2} \frac{3}{2G_0} \alpha +
16 \pi \tilde L_m \right]
\end{equation}
where we have defined $\tilde L_m=L_m/(G_0 G(\phi)^{-1})^2$. It is now possible
to define a new scalar field $\psi$ by

\begin{equation}
d\psi= \sqrt{ \frac{3}{16 \pi G_0} \alpha} \frac{d\phi}{\phi}
\end{equation}
such that,  

\begin{equation}
S_{EF}= \frac{1}{16\pi} \int \sqrt{- \tilde g} \left[ \frac{1}{G_0} \tilde R 
+ 16 \pi (\tilde L_m - \frac12 \psi_{,a} \psi^{,a} ) \right]
\end{equation}
Thus, we recover the Einstein action with a stress energy tensor given by the 
sum of two contributions, the matter and the scalar field ones. It is worth
noting that the scalar field $\psi$ is proportional to the variable $Y$ which was
introduced to solve the problem of FRW models in \cite{TORRES_V}. The recovery of the
Einstein action --and the field equations derived from it-- does, however, have a cost.
That is, a non-independently conserved stress-energy tensor for matter, which in the
Einstein frame behave in agreement with
 
\begin{equation}
\tilde T_{ab}^{\;\;\; ;a}=
\frac12 \sqrt{ \frac{ 16 \pi G_0}{3} \frac{1}{\alpha} } \tilde T
\psi_{;b} \left( \frac{\phi}{G^{-1}} \frac{dG^{-1}}{d\phi} \right)
\end{equation}
So, in any case in which $\tilde T \neq 0$, only the total stress-energy tensor
will be conserved, {\it i.e}. $(\tilde T_{ab} + \tilde T^{\psi}_{ab})^{;a}=0$; 
where $ \tilde T^{\psi}_{ab}$ 
stands for the stress-energy tensor related with the field, 
which is given by

\begin{equation}
\label{psitensor}
\left( \tilde T^{\psi}_{ab} \right)^{;a}=\left( \psi_{,a} \psi_{,b} - \frac12
\psi_{,c} \psi^{,c} \tilde g_{ab} \right).
\end{equation}
Equation (\ref{psitensor}) and cosmological assumptions for the field
show that it behave as a stiff fluid with density and pressure given by:

\begin{equation}
\label{psidensity}
\tilde \rho^{\psi}= \tilde p^{\psi}=\frac12 \left( \frac{d\psi}{d\tilde t} \right)^2
\end{equation}
Due to the exact reproduction of the the Einstein field equations, 
the common results of general relativity will hold good. 
We shall take into account
matter given by a barotropic fluid, particularly in the cases of stiff fluids or
radiation or both. The case of a dust fluid plus radiation in scalar-tensor
theories is analyzed in \cite{TORRES_H} while some cases of imperfect 
fluids in \cite{PIMENTEL2}. Following Raychaudhuri \cite{RAY}, and
considering those models in which the velocity of matter is parallel to the unit normal
to the spatial hypersurfaces (a geodesic time-like vector $t^a$); it is possible to
write the Einstein field equations --the HSTG equations in the Einstein frame--
in the form of a constraint

\begin{equation}
\tilde \theta^2 = 24 \pi G_0 (\tilde \rho + \tilde \rho^\psi) 
+3 \tilde \sigma^2 -\frac32 \;\; ^3 \tilde R
\end{equation}
plus the Raychudhuri equation.

\begin{equation}
\frac{d \tilde \theta}{d \tilde t} +\frac13 \tilde \theta^2 = - 4 \pi G_0 
(\tilde \rho + \tilde \rho^\psi + 3 (\tilde p + \tilde p^\psi))
\end{equation}
where we have introduced the expansion $\theta$, the shear $\sigma$ and the curvature
scalar of
hypersurface of homogeneity $^3 R$; all them, in the Einstein frame.
The transformed quantities are

\begin{equation}
\tilde \rho= \frac{\rho}{\left( G_0 G(\phi)^{-1}\right)^2 }\hspace{0.3 cm} 
\tilde p= \frac{p}{ \left( G_0 G(\phi)^{-1}\right)^2 }  \hspace{0.3 cm} 
\tilde \sigma^2= \frac{\sigma^2}{\left( G_0 G(\phi)^{-1}\right)} \hspace{0.3 cm} 
d\tilde t^2= \left( G_0 G(\phi)^{-1}\right) dt^2
\end{equation}
we also introduce a volume factor $\tilde V =  \left( G_0 G(\phi)^{-1} \right)
^{3/2} V$ with $V$ such that $\theta= dV/Vdt$.

Having in hands both conservation laws, 
for matter in the Jordan frame and for
matter plus field in the Einstein frame, 
it is possible to derive the corresponding
energy densities and pressures. They are:

\begin{equation}
\label{matden}
\tilde \rho= 
\frac{3}{8\pi G_0} \left( \frac{\Gamma}{\tilde V^{4/3}} 
+ \frac{ M G_0 G(\phi)^{-1}}
{ \tilde V^2} \right)
\end{equation}

\begin{equation}
\label{matpre}
\tilde p= 
\frac{3}{8\pi G_0} \left( \frac{\Gamma}{3 \tilde V^{4/3}} 
+ \frac{ M G_0 G(\phi)^{-1}}
{ \tilde V^2} \right)
\end{equation}

\begin{equation}
\label{psiden}
\tilde \rho^{\psi}= \tilde p^{\psi}=
\frac{3}{8\pi G_0} \left( \frac{A^2 - 4 M G_0 G(\phi)^{-1}}
{4 \tilde V^2} \right)
\end{equation}

Here, $\Gamma $ is related with the presence of radiation and $M$
with the presence of a stiff fluid. Both, $\Gamma$
and $M$, are positive constants.
When a stiff fluid is present, it is possible to define
a new field $\chi$, minimally coupled to the metric, such that its energy
density be the sum of the energy density of the scalar $\psi$ and the
matter content. This field $\chi$ can be related with $\phi$ by
\cite{TORRES_V,MWa}

\begin{equation}
\label{rel}
\sqrt {\frac{16 \pi G_0}{3}}\chi(\phi)= A \int \frac{d\tilde t}{\tilde V}=
\pm \int \sqrt \frac{A^2}{ A^2 - 4 M G_0 G(\phi)^{-1}}
\;\;\;  \sqrt {\alpha} \;\;\;\; \frac{d\phi}{\phi}
\end{equation}
We see that, defining a new set of functions $(G,w)$ by

\begin{equation}
\label{rel2}
\alpha_{vac}= \frac{A^2}{ A^2 - 4 M G_0 G(\phi)^{-1}} \alpha ,
\end{equation}
the effect of 
stiff fluid matter is equivalent to a new lagrangian with no matter content.
That change is possible because the stiff fluid modify the dependence of 
$\chi$ with $\phi$, function that in vacuum models is solely accomplish
by $w$ and $G$. That was exactly what happened with BD models, where
only existed $w$.
Also here, the field equations become simplest  with the use of the variable
$X=\left( G_0 G(\phi)^{-1} \right) a^2 = \tilde a^2$
and the conformal time  \cite{TORRES_V}. With this variable
together with $\psi$ given above we can now respond to the two questions outlined
in the begining of this work. 
To do so, let us treat in this formalism and as a first example, 
what happens with
Bianchi I universes.

The metric of Bianchi I models is 

\begin{equation}
\label{metric}ds^2=dt^2 - a_1(t)^2 dx^2  - a_2(t)^2 dy^2
 - a_3(t)^2 dz^2
\end{equation}
Here, the expansion is given in terms of an averaged
scale factor $a^3=a_1a_2a_3=V$ in the form: 
$\theta=3 da/adt$.
The spatial curvature is null and the metric
reduces to the flat FRW one in the case in which
$a_1=a_2=a_3$. The general relativistic 
result for the shear holds in the Einstein
frame; {\it i.e.} 
$\tilde \sigma^2=3 \Sigma^2/4\tilde a^6$, where $\Sigma$ is a constant
and $\tilde a^3=\tilde V$.  
Using the expressions for the energy densities, it is possible
to obtain the constraint 
equation in the variable $X$. For matter given in the form of
non-interacting stiff and radiation fluids it is

\begin{equation}
\label{X'}
X^{\prime 2}= A^2 + \Sigma^2 + 4 \Gamma X
\end{equation}
It has exactly the same form that the equation obtained in \cite{MWa}
and so, it admits the same solution 
(equation (105) of that paper).
Equation (\ref{X'}) is in fact a general relativistic result,
valid here because of the Einstein frame \cite{Ruban}.
It is important to stress that, although the equation and its solution
are the same, the meaning of the variables are different and that it is
now allowed the study of the more general case of lagrangian (\ref{HSTG}).
The solution of (\ref{X'}) shows a shear-dominated evolution at early times
and a radiation-dominated evolution when the averaged scale factor
tends to infinity. Note that these results do not depend neither on
the particular form of $G$ nor of $w$. The specification of $X$, the shear
and each one of the scale factors of the metric --which can be obtained
from the general relativistic
results in the Einstein frame-- describe the full evolution of the system.
To go back to the Jordan frame, we have to obtain $\phi(\eta)$. To do so,
we have to invert (\ref{rel}) and get $\phi(\chi)$ and afterwards, use our
knowledge of $X(\eta)$ to get $\chi(\eta)$. Both operations yields
finally, $\phi(\eta)$. From equation (\ref{rel}) we see that to equal
functional form of $\alpha_{vac}$, equal solution for the field $\phi$
is obtained. Indeed, it means that if a solution for $\phi$ in a 
particular BD gravity with $w=w_{vac}$ is known, it is also a solution
for the set of theories given by $\alpha_{vac}=(2w_{vac}+3)/3$.
Thus, it is possible to define $\alpha_{vac}$ as a classifier of equivalence
sets. Within each one of these sets, the definition of $X$ allows us
to obtain the particular behavior of the averaged scale factor.
The functional form of the metric scale factors does not depend on
$w$ or $G$ in the Einstein frame, while it does in the Jordan frame. 
Is in this frame in which a particular dependence of $G$ on $\phi$
discrime among different behaviors inside a equivalence set.
As was the case in BD gravity, without the specification
of the functions involved in the lagrangian, some conclusions
may arise. In presence of a stiff fluid, a bounce will occur
when

\begin{equation}
\frac{da}{dt}=\frac12 \left( \frac{X'}{X} - G (G^{-1})' \right) =0
\end{equation}
which requires

\begin{equation}
(A^2+\Sigma^2) \alpha=
(A^2-4MG_0G^{-1})\left( \frac{\phi}{G} \frac{dG}{d\phi} \right)^2.
\end{equation}
It may be also seen that, if $G^{-1}$ vanishes faster than $X^3$,
an anisotropic initial singularity in the Einstein frame becomes
isotropic in the Jordan frame. 
Note that the condition for the bounce is incompatible with a 
vanishing $G^{-1}$ and $X$. The possibility of having a finite 
expansion in this situation is analized for BD gravity in \cite{MWa}.

The above analysis implies that the algorithm method developed by
Mimoso and Wands for Brans-Dicke theory is capable to deal also with Bianchi
I models in general hyperextended theories and that the specific
choice of a convenient $w$ allows us to study also all the cases
of
non-minimally coupled gravitation which still retains the positivity of 
$\alpha$, as was the case in BD gravity imposing, accordingly, 
values of $w$ bigger than -3/2. 
The classification scheme 
of FRW models is not broken by anisotropy.

As a matter of fact, one can --at this stage-- observe that
the same applies to others models, such as Bianchi V or III. 
Those models are studied in \cite{Ruban} for General Relativity and in
\cite{MWa} for BD gravity, and the only thing 
necessary to  answer the former questions in this same way is translate that
analysis to the variables $X$, $\psi$ and equations (\ref{rel}) and (\ref{rel2})
of this work.   
In the sake of conciseness we do not do so here.

\vspace{1.2 cm}
This work was partially supported by CONICET.


\begin{thebibliography}{99}

\bibitem{BD}  C. Brans and R.H. Dicke, Phys. Rev. {\bf 124}, 925
(1961)

\bibitem{STG}  P.G. Bergmann, Int. J. Theor. Phys. {\bf 1}, 25 (1968); K.
Nortvedt, Astrophys. J. {\bf 161}, 1059 (1970); R.V. Wagoner, Phys. Rev. 
D {\bf 1},3209 (1970)

\bibitem{INF} R. Fakir and G. Unruh, Phys. Rev D {\bf 41}, 1783 (1990); 
{\it ibid.} {\bf 41}, 1792 (1990)
D. La and P.J. Steinhardt, Phys. Rev. Lett. {\bf 62}, 376
(1989), P.J. Steinhardt and F.S. Ascetta, Phys. Rev. Lett. {\bf %
64}, 2470 (1990)

\bibitem{strings} E. S. Fradkin and A.A. Tseytlin, Nuc. Phys. B {\bf 261}, 1
(1985); C.G. Callan, D. Friedan, E.J. Martinec and M.J. Perry, 
Nuc. Phys. B {\bf 262}, 593 (1985); C. Lovelock, 
Nuc. Phys. B {\bf 273}, 413 (1985)

\bibitem{WILL} C. Will, {\it Theory and Experiment in Gravitational Physics}
(Cambridge University Press, Cambridge, England, 1981)

\bibitem{Nucleo}  A. Serna, R. Dominguez-Tenreiro and G. Yepes, Astrophys.
J. {\bf 391}, 433 (1992); J.A. Casas, J. Garc\'{\i}a-Bellido and M. Quir\'os,
Phys. Lett. B {\bf 94} (1992); D.F. Torres, Phys. Lett. B {\bf 359},
249 (1995); A. Serna and J.M. Alimi, Phys. Rev. D {\bf 53}, 3087 (1996)

\bibitem{Barrow1}  J.D. Barrow, Phys. Rev. D {\bf 47}, 5329 (1993)

\bibitem{Barrow2}  J.D. Barrow and J.P. Mimoso, Phys. Rev. D {\bf 50}, 3746
(1994). The methods developed in this paper were applied in J.D Barrow and
P. Parsons, to appear in Phys. Rev. D {\it gr-qc archive 9607072 }.
See also, A. Serna and J.M. Alimi, Phys. Rev. D {\bf 53}, 3074 (1996)
\bibitem{MW}  J.P. Mimoso and D. Wands, Phys. Rev. D {\bf 51}, 477 (1995)

\bibitem{LiddleWands}  A.R. Liddle and D. Wands, Phys. Rev. D {\bf 45}, 2665
(1992)

\bibitem{TORRES_V} D.F. Torres and H. Vucetich, Phys. Rev. D{\bf 54}, 7373
(1996)

\bibitem{NMCO}  S. Capozziello and R. de Ritis, Phys. Lett. A {\bf 195}, 48
(1994); {\it ibid}. {\bf 177}, 1 (1993) and references therein.

\bibitem{TORRES_S} D.F. Torres, to appear in Phys. Lett. A{\bf 225}, 13 (1997)
{\it gr-qc archive 9610021}


\bibitem{MAHM} For a review see, M.A.H. MacCallum, in {\it General Relativity: An
Einstein Centenary Survey, S. Hawking and W. Israel, eds.} (Cambridge University
Press, Cambridge 1979)

\bibitem{RF}  See for instance, V.A. Ruban and A.M. Finkelstein, 
Lett. Nuovo Cimento {\bf 5}, 289 (1972)


\bibitem{PIMENTEL} L.O. Pimentel, Phys. Lett. B {\bf 246}, 27 (1989)



\bibitem{Chauvet}P. Chauvet and J.L.Cervantes-Cota, 
Phys. Rev. D{\bf 52}, 3416 (1995)


\bibitem{MWa}  J.P. Mimoso and D. Wands, Phys. Rev. D {\bf 52}, 5612 (1995)

\bibitem{Dicke} R.H. Dicke, Phys. Rev. {\bf 125}, 2163 (1962)

\bibitem{Synge} J.L. Synge, {\it Relativity, The General Theory} (Nort Holland
Publishing Company, Amsterdam, 1960), p.318.
 
\bibitem{TORRES_H} D.F. Torres and A. Helmi, Phys. Rev. D {\bf 54}, 6181 (1996)

\bibitem{PIMENTEL2} L.O. Pimentel, Nuovo Cimento B {\bf 109}, 274 (1994)

\bibitem{RAY} A.K. Raychaudhuri, {\it Theoretical Cosmology} (Clarendon Press, Oxford,
1979), p.79.

\bibitem{Ruban} V.A. Ruban, J.E.T.P. {\bf 45}, 629 (1978)


\end{thebibliography}
\end{document}